# A Low-Cost Teapot Effect Experiment for Introductory Physics

Yu-Chen Guo[1,2], Jin-Ming Wang[1], Ying-Xin Li[1]

[1] *Department of Physics, Liaoning Normal University, Dalian, China*

[2] *Department of Physics and Astronomy, University of Pittsburgh, Pittsburgh, USA*

The teapot effect refers to the tendency of a poured liquid to cling to the lip of a container and run down the outside. It is a familiar but physically rich example of flow separation. We present a low-cost experiment for introductory physics laboratories that uses 3D-printed cups, a simple flow regulator, and basic surface treatments to explore this phenomenon in a classroom setting. Students measure the run-off length along the outer wall as an accessible indicator of sticking versus separation and use it to compare the effects of flow velocity and surface wettability. Rather than attempting a full quantitative test of research-level models, the activity is designed to illustrate the inertial–capillary picture of the teapot effect in a form that is experimentally straightforward and pedagogically effective. The experiment connects a familiar everyday observation to fluid inertia, wetting, and interfacial forces in a form that is well suited to introductory instruction.

## 1. Introduction

The teapot effect refers to the familiar tendency of a poured liquid to cling to the lip or spout of a container and run down the outside instead of separating cleanly. Although it is often encountered as a household nuisance, it has long attracted attention as an interesting fluid-dynamical problem.[1] Early analyses emphasized hydrodynamic pressure and streamline bending,[2] while more recent work showed that surface wettability is also a key control parameter, even in the inertial regime. In particular, experiments by Duez *et al.* demonstrated that flow separation depends not only on flow speed and edge geometry, but also strongly on the wetting properties of the solid surface.[3,4]

In introductory physics courses, fluid mechanics is often presented mainly through idealized treatments based on Bernoulli's equation, with limited attention to interfacial effects and real fluid behavior near solid boundaries. Recent teaching literature has also highlighted the value of low-cost, classroom-accessible fluid dynamics activities for introductory students[5]. The teapot effect provides a useful opportunity to broaden that picture. It is easy for students to recognize, visually clear in the laboratory, and well suited to introducing the competition among inertia, wetting, and geometry. In this paper, we adapt these ideas into a low-cost classroom activity in which students use the run-off length along the outer wall as a practical observable for comparing sticking and separation. Our goal is not to reproduce research-grade measurements, but to provide an accessible experiment that illustrates the main inertial–capillary trends in a form suitable for introductory instruction.

## 2. A Simple Picture of the Teapot Effect

A detailed treatment of pouring flows can become mathematically involved, but the basic physical picture is straightforward. When water reaches the lip of a cup, it must either continue outward or bend to follow the surface. Separation is favored when the liquid has enough inertia to maintain its forward motion, while sticking is favored when wetting and streamline bending near the edge promote adhesion to the solid.

A useful dimensionless measure of the relative importance of inertia is the Weber number,

$$We = \rho U^2 e_0 / \gamma$$

where $\rho$ is the fluid density, $U$ is the flow velocity, $e_0$ is the liquid film thickness near the lip, and $\gamma$ is the surface tension.[3] In this classroom version, we use $U$ as the directly controlled experimental parameter and do not attempt a full determination of $We$. Larger values of $We$ correspond to stronger inertial effects and therefore a greater tendency for the flow to separate from the edge. In addition to inertia, previous studies have shown that wettability and edge geometry also influence whether a flowing sheet detaches or remains attached.

In the present classroom version of the experiment, we do not attempt a full quantitative comparison with theoretical predictions for ejection angle or critical Weber number. Instead, we use this inertial–capillary picture to guide student interpretation of two robust trends: increasing flow velocity tends to reduce sticking, and increasing water repellency tends to suppress the teapot effect. This level of interpretation is sufficient for introductory instruction while remaining consistent with the main ideas established in the research literature.

## 3. Experimental Apparatus and Procedure

A practical advantage of this activity is that it can be assembled from inexpensive and readily available materials. The apparatus consists of three primary components: a regulated liquid source, a set of 3D-printed cups, and a simple measurement scale. The experimental apparatus is shown in Figure 1. Three interchangeable cups with different lip radii were fabricated. In the core version of the activity reported here, one cup is used for the main velocity and wettability comparisons, while the others are available for optional extensions on edge geometry.

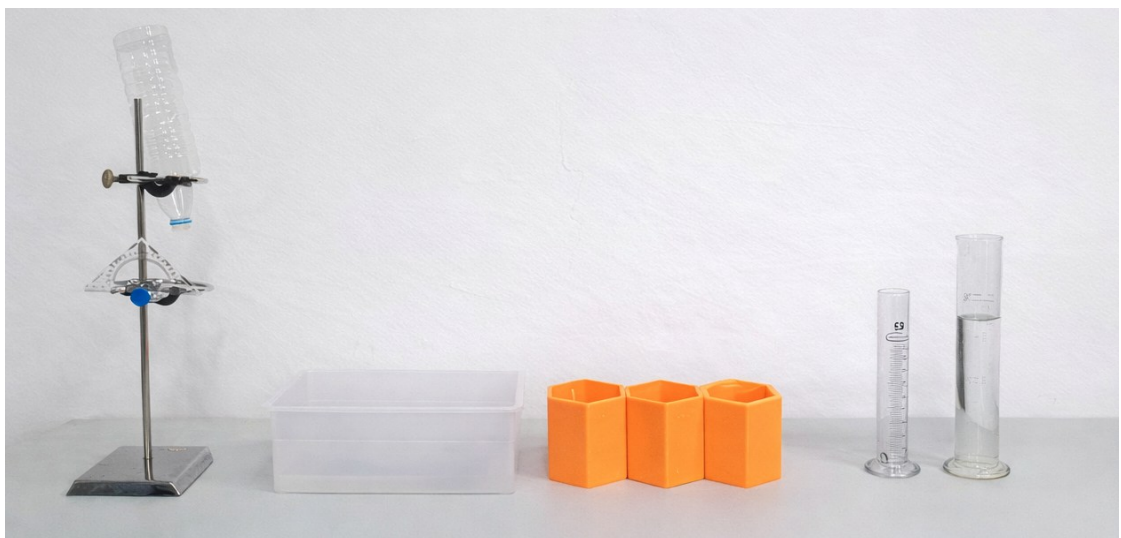

**Fig. 1. Experimental apparatus for the classroom teapot-effect activity, including the regulated liquid source, support stand, collection containers, and the 3D-printed cups used in the experiment.**

### A. Flow Control

Maintaining a steady flow is critical for reproducible $L_{wall}$ measurements. We used a simple constant-head flow system:

1. A 500 mL plastic bottle is placed on an adjustable lab jack. The 3D-printed cup is also secured to the same stand. By adjusting the angle $\theta$ of the cup, the geometry of the flow near the lip can be varied. In the measurements reported here, $\theta$ was fixed unless otherwise noted.

2. A flexible tube connects the bottle to a fine-adjustment needle valve, commonly found in aquarium supplies or medical IV sets.

3. The flow is directed into the 3D-printed cup. Before each trial, the fluid flux $Q$ is measured by collecting water in a graduated cylinder for $\Delta t = 10$ s. The average velocity $U$ at the lip is then calculated using the formula $U = Q/A$, where $A$ is the cross-sectional area of the stream, as determined using a digital caliper.

**B. Measuring $L_{wall}$**

A waterproof millimeter scale is adhered vertically to the outer wall of the cup. For each flow velocity, students record the run-off length $L_{wall}$, defined as the vertical distance from the lip to the point where the water stream visibly detaches from the solid surface.

**C. Surface Modification Protocol**

To explore interfacial effects, students first take measurements using the untreated PLA surface. They then repeat the measurements after applying a thin, uniform paraffin-wax coating to the lip, which makes the surface more water repellent. This allows for a direct comparison of how surface wettability changes the fluid's tendency to remain attached or detach from the edge.

## 4. Data Analysis and Discussion

The laboratory activity guides students from raw observations to a qualitative interpretation based on the inertial–capillary picture of the teapot effect. Measurements of the run-off length serve as the primary classroom probe of sticking versus separation for a given geometry and surface condition.

**A. Effect of flow velocity**

As students increase the flow velocity, they observe a clear reduction in the run-off length $L_{wall}$, as shown in Figure 2. At lower velocities, the water remains attached to the outer wall over a relatively long distance, indicating pronounced sticking. As the velocity increases, the stream separates more readily from the edge and the measured run-off length becomes shorter. At the highest flow setting used here, the run-off length becomes very small, indicating near-clean ejection with only minimal residual contact along the wall.

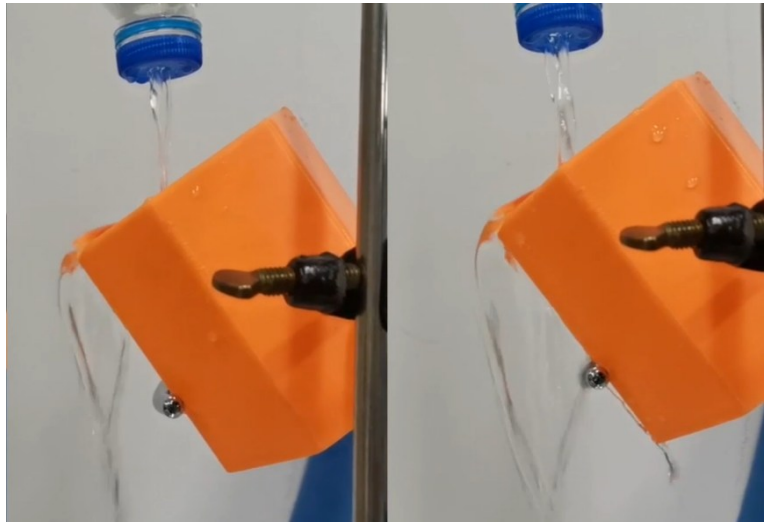

**Fig. 2. Representative photos showing the effect of flow velocity on the teapot effect. At a higher flow velocity (left), the stream separates more readily from the edge. This results in a shorter run-off length and cleaner ejection. At a lower velocity (right), the liquid clings to the lip and runs further down the outside of the cup.**

This trend is consistent with the basic inertial–capillary picture introduced above. A faster stream carries greater forward momentum and is therefore less easily deflected into the curved path required for sticking. In classroom discussion, this result provides a natural way to connect a familiar observation to the idea that flow separation is governed by the competition between inertia and adhesion at a wetted boundary.

Students can summarize these data in a simple plot of $L_\text{wall}$ versus flow velocity $U$, as shown in Fig.3. Under the fixed conditions $\theta = 30°$, $r_i = 0.6$, and untreated PLA, the decrease of $L_\text{wall}$ with increasing $U$ provides a simple quantitative summary of reduced sticking. Each flow setting was measured three times, and the spread of the resulting values is shown as the error bar in Fig. 3. Although the measurements are modest in precision, the qualitative trend is robust and well suited to introductory laboratory work.

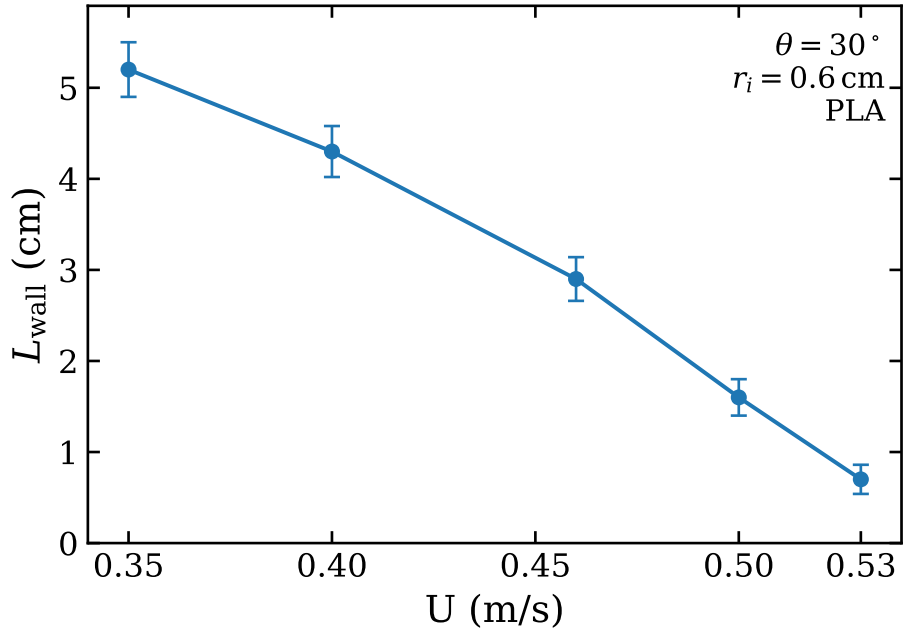

**Fig. 3.** Average run-off length $L_{wall}$ as a function of flow velocity for a fixed cup geometry and surface condition. Error bars indicate the spread from repeated trials. The decrease of $L_{wall}$ with increasing flow velocity reflects the reduced tendency of the liquid to remain attached to the outer wall.

## B. Effect of surface wettability

For a fixed flow velocity and cup geometry, changing the surface condition produces a clear change in the run-off length. In our classroom version, students compare the as-printed PLA surface with the same edge after coating it with a thin layer of paraffin wax. Under otherwise similar conditions, the waxed surface consistently produces a shorter run-off length than the bare PLA surface, indicating weaker sticking and easier separation.

This comparison is especially useful in instruction because it shows that the outcome is not determined by flow velocity alone. Two flows with similar apparent strength can behave differently simply because the liquid interacts differently with the solid surface. In this way, the experiment helps students recognize that fluid motion near a surface depends not only on bulk inertia but also on interfacial properties such as wetting.

The result is consistent with the inertial–capillary picture discussed earlier. A more wetting surface favors adhesion of the liquid near the lip, while a more water-repellent surface reduces that tendency and allows the flow to detach more readily. For introductory students, this provides a useful extension beyond ideal-fluid treatments and opens discussion of contact angle, wetting, and the role of boundary conditions in real flows.

In practice, the wettability study works well as a side-by-side comparison. Students can compare the untreated and wax-coated versions of the same lip under the same nominal flow condition and then rank the surfaces according to the measured run-off length. This part of the activity usually leads to productive discussion about contact angle, wetting, and why a more water-repellent surface can reduce the tendency of the stream to wrap around the lip.

### C. Suggested presentation of student data

For introductory courses, we have found it useful to keep the data analysis simple. One option is to have each group repeat every trial three times, calculate an average run-off length, and display the results with hand-drawn or spreadsheet-generated graphs. A plot of $L_{\text{wall}}$ against flow velocity can be paired with a bar chart comparing different surface conditions. These figures are sufficient to reveal the main trends without requiring advanced image analysis or specialized instrumentation.

If instructors wish to extend the activity, students may also record slow-motion videos with a smartphone and compare the visible shape of the stream near the lip. Such recordings can help students distinguish between partial sticking and clean ejection, but they are not essential for the core version of the experiment. The central learning goal is already achieved when students can connect changes in flow velocity and surface wettability to systematic changes in the run-off length.

## 5. Pedagogical Reflections and Conclusion

The teapot effect offers a useful example of how a familiar phenomenon can be turned into a simple quantitative classroom activity with clear physical interpretation. With inexpensive materials and simple measurements, students can investigate how flow separation depends on two accessible control parameters: the velocity of the liquid and the wettability of the surface. The resulting observations provide a concrete setting for discussing inertia, capillary adhesion, and boundary effects in fluid motion.

An important practical feature of this activity is that it does not require students to extract a precise ejection angle or carry out detailed fluid-dynamical modeling. Instead, the run-off length along the outer wall serves as an observable that is easy to define, easy to measure, and clearly related to the strength of sticking. This makes the experiment well suited to introductory instruction, where clarity and reproducibility are often more important than formal completeness.

The activity is also flexible. In its simplest form, it can be completed in a single lab period as a guided comparison of flow velocities and surface treatments. In a more advanced version, instructors may invite students to explore additional variables such as lip sharpness or edge curvature, or to compare the observations with a simple inertial–capillary interpretation. These extensions make it possible to adapt the same apparatus to different course levels without changing its central idea.

This experiment also illustrates a broader teaching principle: students often engage more readily with formal ideas when those ideas are connected to familiar experience. The teapot effect is immediately recognizable, visually striking, and experimentally accessible. For these reasons, it provides an effective bridge between everyday observation and introductory fluid dynamics.

**Acknowledgments**

This work was supported by the Liaoning Provincial Education Sciences Planning Program under Grant No. JG24DB314.